\documentclass[aps,prl,twocolumn,showpacs,superscriptaddress,groupedaddress]{revtex4}  
\usepackage{graphicx}  
\usepackage{dcolumn}   
\usepackage{bm}        
\usepackage{amssymb}   
\usepackage[latin1]{inputenc}
\usepackage{mathrsfs}

\begin{document}

\title{Exact density-functional potentials for time-dependent quasiparticles}
\author{J.~D.~Ramsden} \affiliation{Department of Physics, University of York, Heslington, York YO10 5DD, United Kingdom} \affiliation{European Theoretical Spectroscopy Facility (ETSF)}
\author{R.~W.~Godby} \affiliation{Department of Physics, University of York, Heslington, York YO10 5DD, United Kingdom} \affiliation{European Theoretical Spectroscopy Facility (ETSF)}
\date{\today}

\begin{abstract}
We calculate the exact Kohn-Sham potential that describes, within time-dependent density-functional theory, the propagation of an electron quasiparticle wavepacket of non-zero crystal momentum added to a ground-state model semiconductor. The potential is observed to have a highly nonlocal functional dependence on the charge density, in both space and time, giving rise to features entirely lacking in local or adiabatic approximations. The dependence of the non-equilibrium part of the Kohn-Sham electric field
on the local current and charge density is identified as a key element of the correct Kohn-Sham functional.
\end{abstract}

\maketitle

Density functional theory (DFT) in the local density approximation (LDA) has proven highly effective in the ground-state calculation of electronic structures and energies of a diverse range of systems \cite{ref1}. It was natural therefore to extend the approximation to the adiabatic LDA (ALDA) within a time-dependent density-functional-theory (TDDFT) description of excited matter. However, while the ALDA is effective in certain classes of excited systems \cite{ref2}, it is shown to break down in a large number of physical scenarios of universal interest, for instance the absorption spectra of semiconductors and insulators \cite{ref3}, the polarization of systems subject to an external electric field \cite{ref4,ref5}, and the prediction of correlation energies \cite{ref6} and double excitation spectra \cite{ref7}.

One important class of non-equilibrium problems for which accurate TDDFT functionals are needed is the quantum transport of charge through matter. Historically, quantum treatments of electronic transport reduce the current response for an applied potential across a device to a transmission problem, and employ non-interacting electrons. More accurate calculations make use of effective potential fields including Hartree, exchange and approximate exchange-correlation terms. A natural step, then, is a Kohn-Sham (KS) TDDFT approach in which the exact potential is a functional of the initial state and the time-dependent density. The need to better understand Kohn-Sham functionals, including those depending explicitly on the current density, beyond the adiabatic and exchange-only regimes in order to describe quantum transport has been previously noted \cite{ref8}. Maitra {\it et al.}  \cite{ref9} have also highlighted the importance of including the initial-state dependence, absent in adiabatic approximations, of 
the time-dependent Kohn-Sham potential in the desired functional.

To advise the construction of appropriate functionals, we believe it is important to study the exact Kohn-Sham potential itself. The reverse-engineering of ground-state KS potentials via direct numerical optimization of characteristic parameters has provided insight into the nature of nonlocal functional dependence of, for instance, polarized materials \cite{ref4,ref5}, while in the time-dependent regime there have been calculations of exact time-dependent exchange-correlation potentials and kernels for non-equilibrium systems such as doubly-excited helium \cite{ref10} and small Hubbard chains subject to time-dependent electric fields \cite{ref11}.

In this paper we introduce a self-consistent algorithm for reverse-engineering the exact potential from known time-dependent charge and current densities. To this end, we consider systems which may be solved exactly by other means, and yet which contain important physical aspects of quantum transport. Our model system is an electron propagating through a infinite, one-dimensional semiconducting wire. The addition of a single electron to a ground-state device -- in this instance the semiconductor wire -- is the domain of quasiparticle theory, in which the wavefunction and energies are given by the Schr\"odinger-like equation:
\begin{equation}
\left[ -\frac{1}{2}\nabla^2 + v_{ext} + v_H - E \right] \psi (\textbf{r}) + \int \Sigma ( \textbf{r}, \textbf{r}', E) \psi (\textbf{r}') d\textbf{r}' = 0
\end{equation}
where $v_{ext}$ is the external potential, $v_H$ the Hartree potential, $E$ the quasiparticle energy and $\Sigma$ the nonlocal, energy-dependent self-energy operator.

To focus on the physics arising from the nonlocality of the self-energy operator, we take the net local potential of the ground-state $V_{ext}+V_H$ to be zero and employ a model operator that is nonlocal but real and energy-independent \cite{ref4,ref12}:
\begin{equation}
\Sigma(x, x') = \frac{f(x) + f(x')}{2}g(\left| x - x' \right| )
\end{equation}
where $g(x) = \text{exp}\left( -(x/w)^2 \right)/\sqrt{\pi}w$ introduces the nonlocality of the self-energy, and $f(x) = -F_0\left[1 - \text{cos}(2\pi x/a)\right]$ is the local, periodic potential to which $\Sigma$ would reduce if $w$ were small. We choose our operator to approximate that of silicon, with a self-energy amplitude of $F_0 = 4.1$ eV, lattice spacing $a=4$ a.u. and nonlocal range $w=2$ a.u.
We model the infinite wire as a chain of "atoms" each contributing one spinless electron to the ground-state.

\begin{figure}
\includegraphics[scale=0.7,trim=60 50 400 195]{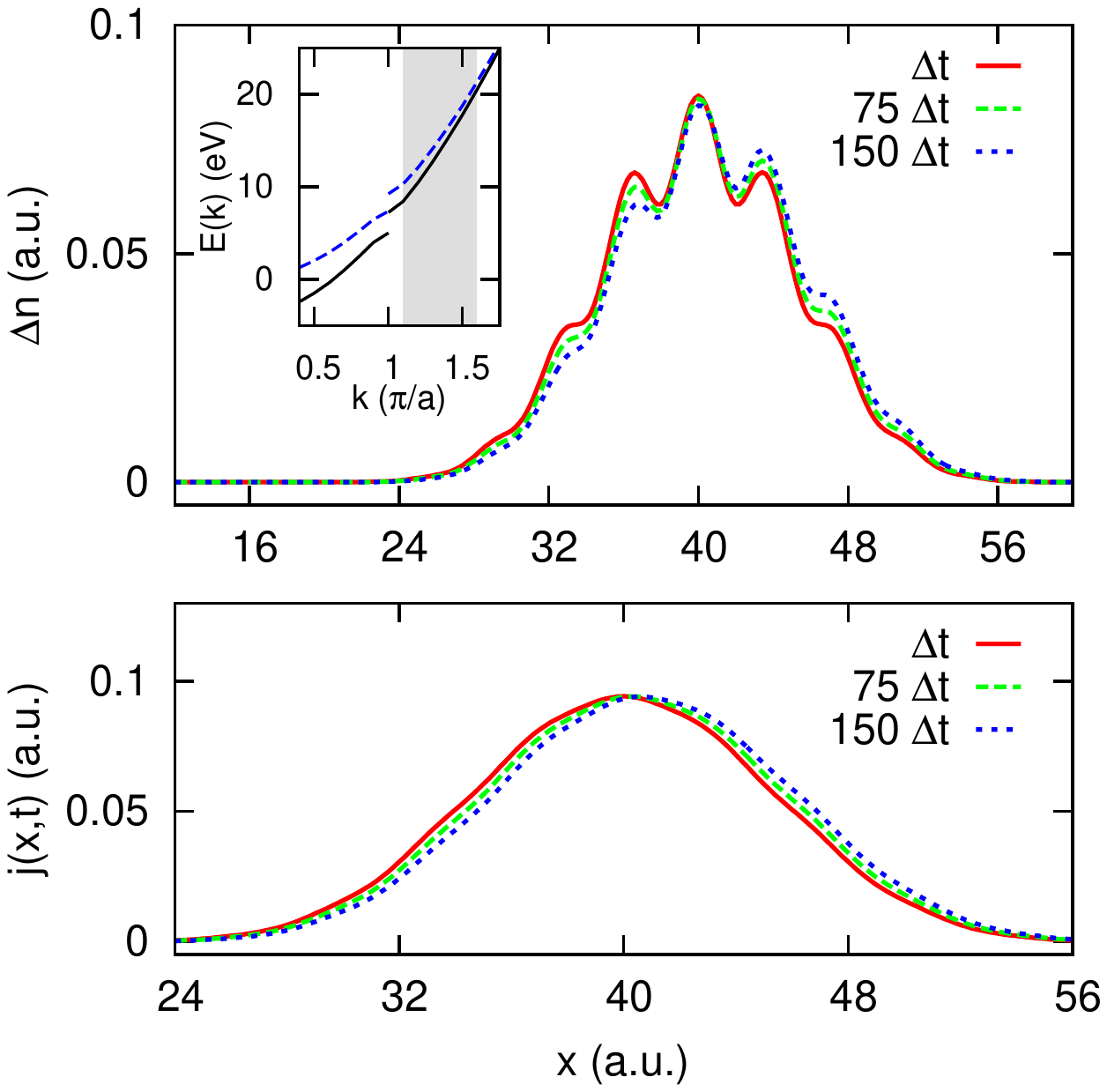}
\caption{\label{fig:rgfig1} (Color online) The charge density additional to the ground state (top) and the current density (bottom) of both the model and Kohn-Sham systems at 1 time-step (solid red), 75 time-steps (dashed green) and 150 time-steps (dotted blue) of $\Delta t = 4 \times 10^{-3}$ a.u. The quasiparticle (solid black) and Kohn-Sham (dashed blue) band structures are inset, with the gray shaded region marking the range of pure quasiparticle Bloch states included in the electron wavepacket.}
\end{figure}

The Kohn-Sham ground state is then described by a scalar potential which reproduces the model ground-state density exactly, which, since it is periodic, may be found by direct numerical optimization of its Fourier components. The resultant KS band structure (Fig. \ref{fig:rgfig1}) differs from that of the nonlocal model, with a band gap that falls short of the quasiparticle gap by 0.31  eV \cite{ref4}. The difference in band structure gives rise to a difference in group velocity $\left( v_g = dE/dk \right)$ between the Kohn-Sham and quasiparticle eigenstates, which would propagate a wavepacket incorrectly and thus necessitates a time-dependent component in the KS potential.

In contrast to typical approaches to nonequilibrium systems, we will prepare the system ``in flight'', taking an arbitrary instant of the quasiparticle propagation as our $t=0$. The addition of a \textit{pure} quasiparticle with a single Bloch wavevector to a ground-state semiconducting wire places the device in the steady-state regime; in this paper we focus on an explicitly time-dependent system: the addition of a quasiparticle \textit{wavepacket} with a non-zero crystal momentum that is sufficiently well-defined that the quasiparticle approximation can be applied. Thus we construct the quasiparticle wavepacket as a weighted sum over a narrow range of solutions to the quasiparticle equation with a momentum expectation value $\left< p_{QP} \right> = 0.65 \times 2\pi/a$.

Because of the initial-state-dependence of the exact time-dependent potential, it is common to prepare the Kohn-Sham system in its ground state, where the exact potential is a unique functional of the density, and, at some arbitrary time, induce an excitation with a chosen time-dependent potential or else excite it toward some desired state or density via optimal control \cite{ref13}. However, the KS wavefunction at some intermediate time between its preparation in the ground state and a later excited state of interest implies a \textit{prehistory} without further explicit functional dependence on the time-dependent density that preceded it \cite{ref9}. Thus we may choose an initial KS state that is in-flight \cite{ref14}, as for the model system, first with the condition that it yields the same initial charge density, and then by using the transformation
\begin{equation}
\psi(x,0) \leftarrow e^{i\theta(x,0)}\psi(x,0)
\end{equation}
and optimizing the real field $\theta(x)$ to yield the correct current density.

The quasiparticle wavepacket, upon being introduced into the supercell, propagates without the necessity of a potential difference across the wire. (The Crank-Nicolson method \cite{ref15} is employed to solve the time-dependent Kohn-Sham and quasiparticle equations explicitly.) Figure \ref{fig:rgfig1} shows snapshots of the time evolution of the quasiparticle charge density $n_{QP}(x,t) = n(x,t) - n_{GS}(x)$ and current density $j_{QP}(x,t)$ as it propagates a short distance along the wire. For reasons that will become clear below, we restrict ourselves in this paper to small displacements of the wavepacket.

\begin{figure}
  \includegraphics[trim=130 285 130 285, clip=true, scale=0.95]{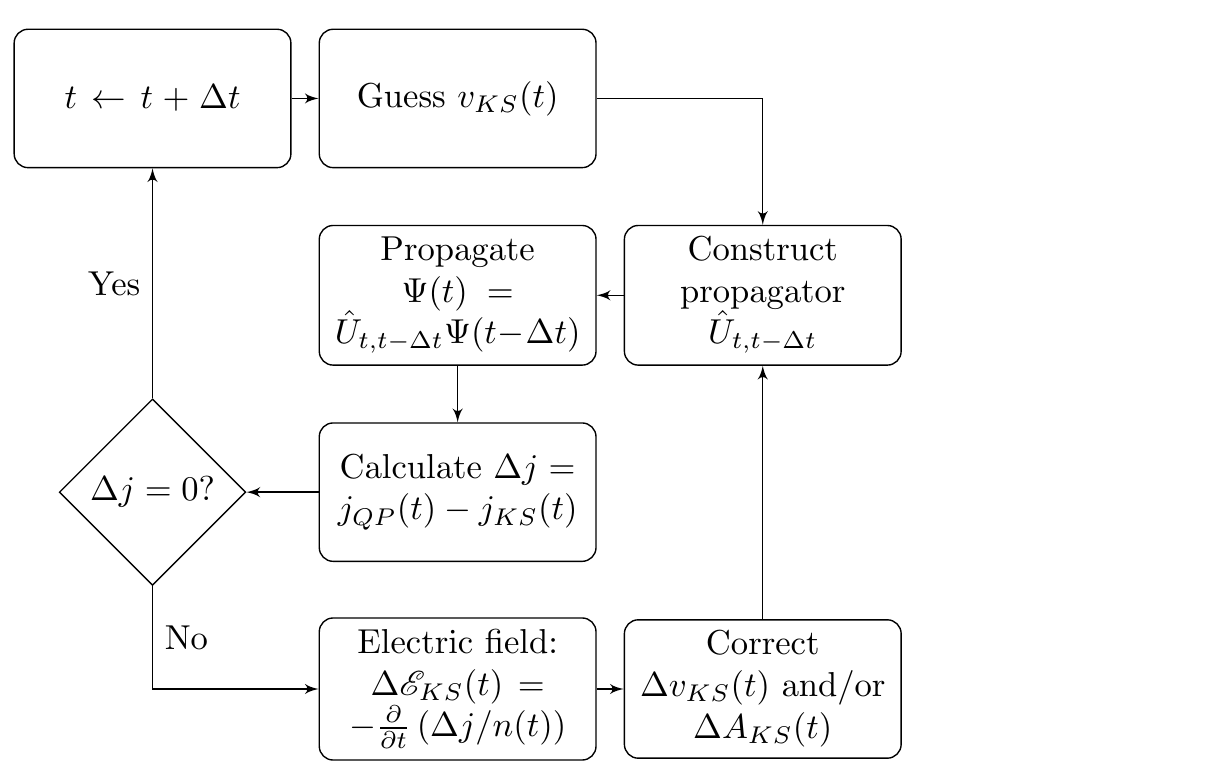}
\caption{\label{fig:rgfig2} \textbf{The reverse-engineering algorithm.} A self-consistent calculation of the exact (in the limit of $\Delta t \rightarrow 0$) time-dependent Kohn-Sham potential which is repeated for every time-step in the quasiparticle system history. The algorithm makes incremental corrections to the Kohn-Sham current density, with the correct charge density arising due to the continuity equation. Here, $v_{KS}$ is the Kohn-Sham potential, $\Psi$ is the Kohn-Sham Slater determinant, $\Delta j$ is the current density error, and $\hat{U}_{t,t-\Delta t}$ is the unitary evolution operator which time-evolves $\Psi$ from time $t-\Delta t$ to time $t$.}
\end{figure}

The algorithm outlined in Figure \ref{fig:rgfig2} is then invoked at each time-step to calculate the exact Kohn-Sham potential. At the heart of the algorithm lies the observation that, if the instantaneous current density for a Kohn-Sham system in the adiabatic limit is incorrect by an amount $\Delta j = j(x,t) - j_{KS}(x,t)$, one can alter the effective \textit{vector} potential $A_{KS}(x,t)$ to yield a better current density $j_{KS}(x,t) = j_0[\Psi_{KS}] + A_{KS}(x,t)n(x,t)$ (where $j_0[\Psi]$ is the standard single-particle current density) without  affecting the instantaneous charge density $n_{KS}(x,t)$, since the adiabatic vector potential performs a transformation of the Kohn-Sham system of the form of Eq. 3. The algorithm demands that the time-step employed in the time-evolution of the Kohn-Sham system must be small enough such that, if the correct KS current density is obtained for consecutive time-steps, the change in charge density is correct by necessity of the continuity equation $\partial n / \
partial t+\nabla \cdot \textbf{j} = 0$. Since the history of the KS density must match that of the quasiparticle system, the same time-step (here $4\times 10^{-3}$ a.u.) is used for both.

The vector potential, while useful procedurally (Fig. \ref{fig:rgfig2}), is known to generally have a nonlocal (in time) functional dependence on the charge density \cite{ref16}. Of more universal relevance is the Kohn-Sham \textit{electric field} which may be defined as $\mathscr{E}_{KS}(x,t) = -\partial_t A_{KS}(x,t) - \partial_x v_{KS}(x,t)$ which may then be implemented in the Kohn-Sham equations via Kohn-Sham vector and/or scalar potentials as appropriate.

\begin{figure}
  \includegraphics[trim = 130 320 225 120,clip=true]{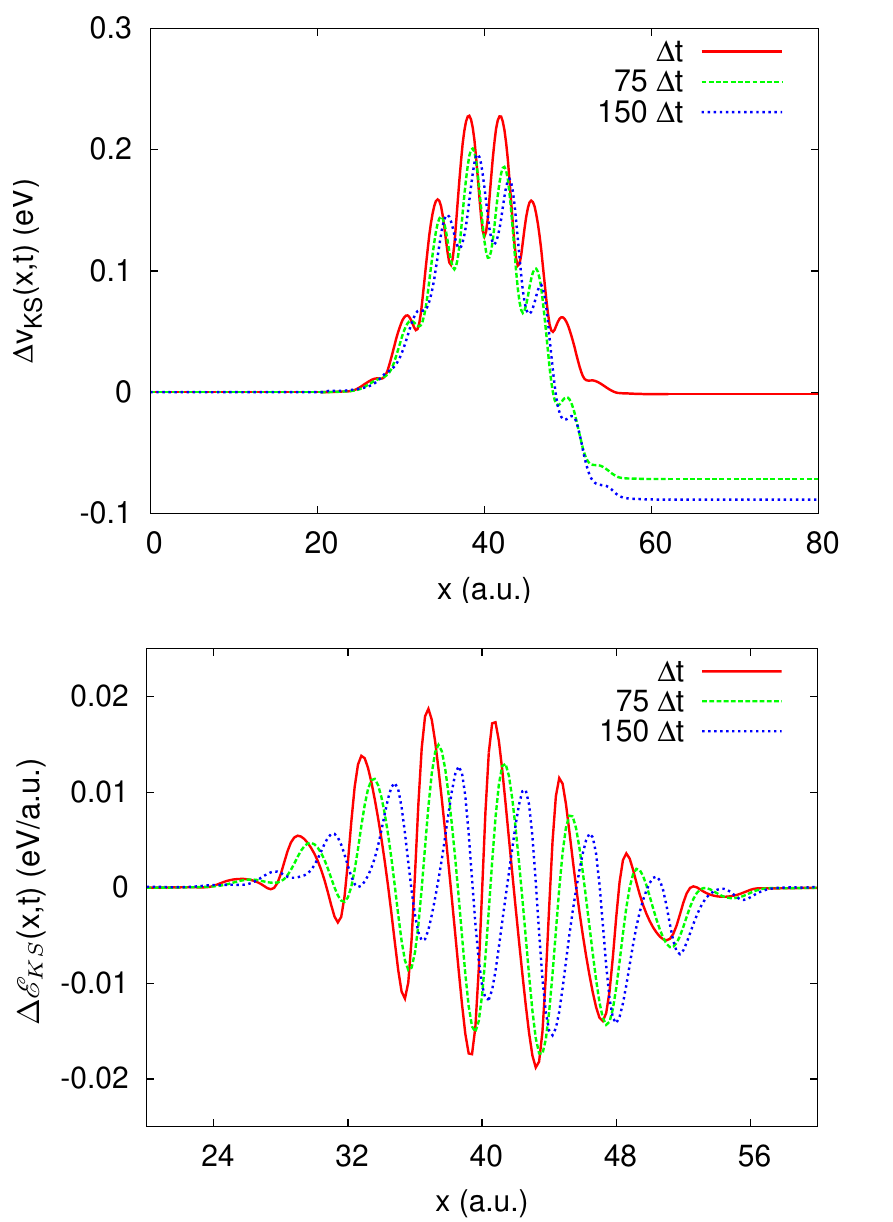}
\caption{\label{fig:rgfig3} (Color online) The additional (relative to the ground state) time-dependent Kohn-Sham potential $\Delta v_{KS}$ (top) and electric field $\Delta \mathscr{E}_{KS} = -\partial v_{KS}/\partial x$ (bottom) after 1 time-step (solid red), 75 time-steps (dashed green) and 150 time-steps (dotted blue) of $\Delta t = 4 \times 10^{-3} a.u.$ The gradient is localized and more slowly-varying, making it more amenable to approximation than the potential itself.}
\end{figure}

Snapshots of the exact additional Kohn-Sham scalar potential for the quasiparticle wavepacket system are shown in Figure \ref{fig:rgfig3}. We observe first that the exact potential is not periodic in space. For this reason, and noting that the quasiparticle is itself finite in space, only a finite section of the Kohn-Sham infinite wire was modelled, specifically a chain of 20 ``atoms" giving a total electron number of 21. We observe that, at all times, the charge densities on each side of the wavepacket region are that of the ground-state wire, while the potentials differ by a time-dependent constant. This ultranonlocal nature of the potential difference negates the possibility of the exact potential being approximated by a local functional of the charge, or indeed current, density. Such potential steps in the exact time-dependent Kohn-Sham potential have been observed before in electron ionization calculations \cite{ref17} and have been linked to the exchange-correlation derivative discontinuity \cite{ref18}
 as the electron number of the system passes through an integer.

Our results differ from previous calculations, however, in three respects. First, although the electron number in a finite section of our system will gradually increase from $N$ to $N+1$ and back again as the wavepacket passes through it, the electron number of the system as a whole remains constant. Second, the potential step of \cite{ref17} takes the form of a plateau which expands from the central region as the ionized electron escapes the system, while, in our system, the ends of the wire are immediately placed in a potential difference due to the localised electric field of the added electron. Third, this potential difference is time-dependent, thus a section of the wire the wavepacket has already passed through will continue to react to electron motion in principle infinitely far away.

By comparing Figures \ref{fig:rgfig1} and \ref{fig:rgfig3}, which feature snapshots at the same time intervals, one observes that this nonlocal aspect of the potential time-varies much more rapidly than the wavepacket charge or current density. Semi-classically, one can understand the necessity that such an electric field be time-varying: an electron subject to a constant electric field will acquire a constant acceleration. Thus a static potential difference cannot reproduce the steady progression of the actual quasiparticle, and a time-varying field must be present.

The additional KS potential (Figure \ref{fig:rgfig3} also exhibits a small barrier, localized with the quasiparticle, and modulated by a periodic function with the periodicity of the crystal lattice. The barrier acts as a source of resistance to the current density for the Kohn-Sham electrons while the periodic part of the additional potential acts to `tune' the local band structure, and hence the group velocity, of the Kohn-Sham electrons and, in conjunction with the rest of the barrier, ensures that the shape of the charge density evolves correctly.

It is the combination of these three components of the exact time-dependent potential that provides the mechanism for yielding the correct current density. In contrast to the quasiparticle description, in which all of the current is carried by the added quasiparticle, a small but significant proportion (about 5\%) of the Kohn-Sham current density is carried by the valence-band electrons which are subject to the same additional potential. The dependence of the KS potential on the semilocal \textit{current} density of the system introduces a history-dependence not reflected by the ALDA or other adiabatic functionals of the charge or current density. The current density error $\Delta j$ of a putative Kohn-Sham system that reproduces the exact initial charge, but not current, density remains present in the KS potential thereafter via the localized KS electric field which propagates alongside the wavepacket.

The self-consistent method for calculating the exact Kohn-Sham electric field, while demonstrated for a one-dimensional system in Figure \ref{fig:rgfig2}, is itself independent of dimensionality. While it is trivial to include such a field in the time-dependent Kohn-Sham equations in one dimension, one could generalize to two- and three-dimensional systems with the inclusion of a KS vector potential, and an associated KS magnetic field, as well as a scalar potential. Moreover, the method is, in principle, both universal and exact in the continuous limit, providing exact time-dependent Kohn-Sham potentials in the study of new, better functionals.

The exact TDDFT Kohn-Sham potential for localized quasiparticle systems demonstrates the failure of local and adiabatic functionals such as the ALDA to describe them. A component of the exact potential has a functional dependence on the charge and current density that is highly nonlocal in both space and time, and is a crucial component of the complementary mechanisms that the exact potential employs. However, the inclusion of such effects in explicit scalar-potential functionals is intrinsically problematic. An equivalent problem arises in the velocity gauge, with the presence of the nonadiabaticity of the exact Kohn-Sham vector potential functional \cite{ref16}. The explicit inclusion of a Kohn-Sham electric field functional, however, may provide an additional ingredient that allows nonlocal components of the potential to be implemented via calculation of a local or semilocal functional of the charge and current density.

We thank Peter Bokes for fruitful discussions, and acknowledge funding from the EPSRC.


\begin{thebibliography}{99}

  \bibitem{ref1}
R.~O.~Jones and O.~Gunnarsson, 
Rev. Mod. Phys. {\bf 61}, 689 (1989).

  \bibitem{ref2}
M.~Petersilka, U.~J.~Gossmann and E.~K.~U.~Gross,
Phys. Rev. Lett. {\bf 76}, 1212 (1996).

  \bibitem{ref3}
G.~Onida, L.~Reining and A.~Rubio,
Rev. Mod. Phys. {\bf 74}, 601 (2002).

  \bibitem{ref4}
R.~W.~Godby and L.~J.~Sham,
Phys. Rev. B {\bf 49}, 1849 (1994).

  \bibitem{ref5}
X.~Gonze, Ph.~Ghosez and R.~W.~Godby,
Phys. Rev. Lett. {\bf 74} 4035 (1995).

  \bibitem{ref6}
M.~Lein, E.~K.~U.~Gross and J.~P.~Perdew,
Phys. Rev. B. {\bf 61}, 13431 (2000).

  \bibitem{ref7}
N.~T.~Maitra, F.~Zhang, R.~J.~Cave and K.~Burke,
J. Chem. Phys. {\bf 120}, 5932 (2004).

  \bibitem{ref8}
S.~Kurth, G.~Stefanucci, C.~O.~Almbladh, A.~Rubio and E.~K.~U.~Gross,
Phys. Rev. B {\bf 72}, 035308 (2005);
M.~Koentopp, C.~Chang, K.~Burke and R.~Car
J. Phys. Condens. Matter {\bf 20} 083203 (2008).

  \bibitem{ref9}
N.~T.~Maitra, K.~Burke and C.~Woodward,
Phys. Rev. Lett. {\bf 89}, 023002 (2002).

  \bibitem{ref10}
M.~Thiele, E.~K.~U.~Gross and S.~K\"{u}mmel,
Phys. Rev. Lett. {\bf 100}, 153004 (2008).

  \bibitem{ref11}
C.~Verdozzi,
Phys. Rev. Lett. {\bf 101}, 166401 (2008).

  \bibitem{ref12}
R.~W.~Godby, M.~Schl\"{u}ter and L.~J.~Sham,
Phys. Rev. B {\bf 37}, 10159 (1988).

  \bibitem{ref13}
W.~Zhu, J.~Botina and H.~Rabitz,
J. Chem. Phys. {\bf 108}, 1953 (1998).

  \bibitem{ref14}
In this case, we employ the following physical considerations to calculate the initial Kohn-Sham system: first, that subject to the ground-state Kohn-Sham potential only, the system has the correct crystal momentum; second, that subject to an additional Kohn-Sham potential, the system acquires the correct charge density; third and finally, to yield the correct current density and implement nonadiabatic behavior, the wavefunctions are subject to the transformation described in Eq. 3.

  \bibitem{ref15}
J.~Crank and P.~Nicolson,
Proc. Camb. Phil. Soc. {\bf 43}, 50 (1947).

  \bibitem{ref16}
G.~Vignale and W.~Kohn,
Phys. Rev. Lett. {\bf 77}, 2037 (1996).

\bibitem{ref17}
M.~Lein and S.~K\"{u}mmel,
Phys. Rev. Lett. {\bf 94}, 143003 (2005).
D.~Vieira, K.~Cappelle and C.~Ullrich,
Phys. Chem. Chem. Phys. {\bf 11}, 4647 (2009).

\bibitem{ref18}
L.~J.~Sham and M.~Schl\"{u}ter,
Phys. Rev. Lett. {\bf 51}, 1888 (1983).

\end{thebibliography}
\end{document}